# EFFECT OF DUST GRAIN SPUTTERING ON SELF-GRAVITATIONAL DUSTY PLASMAS

**R. Annou[1,3], R. Bharuthram[1] and P. K. Shukla[2]**
**[1] University of Kwazulu-Natal, Durban, South-Africa**
**[2] Fakultät fûr Physik and Astronomie, Ruhr-Universitat Bochum, D-44780 Bochum, Germany**
**[3] Permanent address: faculty of Physics, USTHB, Algiers. Algeria**

**Abstract**

It is shown that the grain mass fluctuation, which is due to the grain sputtering by the attracted ions, may be at the root of a new instability in infinite, unmagnetized and homogeneous dusty plasmas. The growth rate is determined in the framework of two plasma models, viz., a cold and electron depleted plasma, and a plasma where electrons and ions are taken Boltzmannian. This instability may affect the Jeans as well as the agglomeration instabilities.

Dusty plasmas are known to be encountered in numerous astrophysical situations such as protoplanetary nebulae, interstellar media, molecular clouds[1]..etc. As a matter of fact, the dust component is essential for many processes, e.g., star formation and planetary structures. Indeed, when the grain mass is important, a competition between the electromagnetic as well as the gravitational forces arises that leads to the distinction of different regimes[2], viz., when the gravitational force prevails, large structures are formed due to molecular clouds collapse, whereas if the competing forces are balancing one another, a critical limit is set on the Jeans instability and structures like spokes in Saturn's rings may be explained. This balance is affected by many factors such as, dust size distribution, magnetic field, collisions of dust with ions/neutrals[3] and dust grain charge fluctuations.

In this note we address a new factor that is the erosion of the grain by the ions falling onto the surface[4]. We confine ourselves however, to the sputtering mechanism where the ions strike the grain surface and transfer momentum to the grain atoms that are ejected consequently. The grain mass becomes a dynamical variable and dust mass fluctuations are introduced in the stability analysis.

Let us consider an electron depleted dusty plasma consisting of ions and charged grains, the evolution of which is governed by the following equations (c.f. Ref.[5]),

$$\frac{\partial n_i}{\partial t} + \frac{\partial n_i v_i}{\partial x} = 0, \tag{1}$$

$$\frac{\partial v_i}{\partial t} + v_i \frac{\partial v_i}{\partial x} + \frac{e}{m_i}\frac{\partial \boldsymbol{f}}{\partial x} = -\frac{\partial \boldsymbol{y}}{\partial x} - \frac{v_{thi}^2}{n_i}\frac{\partial n_i}{\partial x}, \tag{2}$$

where $v_{thi}$ is the ion thermal velocity.

In our study, we consider a grain eroded by the ions striking it. The loss in the mass is modelled by a sink term in the continuity equation. The corresponding loss in momentum is in average null for the sputtering process is taken isotropic. The dust continuity and momentum equations are cast as follows,

$$\frac{\partial nm}{\partial t} + \frac{\partial nmv}{\partial x} = -\lambda\, nm^{2/3}, \tag{3}$$

$$\frac{\partial mv}{\partial t} + v\frac{\partial mv}{\partial x} - eZ\frac{\partial \phi}{\partial x} = -m\frac{\partial \psi}{\partial x} - m\frac{v_{thd}^2}{n}\frac{\partial n}{\partial x}, \tag{4}$$

where, $\lambda = \sqrt{\frac{8\pi T_i}{m_i}}\, m_t n_i Y_s (1+x_0)\exp(-x_0)\left(\frac{3}{4\pi\rho}\right)^{2/3}$, $m_t$ is the mass of the atoms composing the grain, $Y_s$ is the sputtering yield that gives an indication of the number of atoms composing the grain released per striking particle, and depends on the nature and energy of the incident particle as well as the nature of the surface of the dust grain, $\rho$ is the grain mass density, $x_0 = U_0/\eta T_i$ where $U_0$ is the binding energy of an atom at the surface layer of the solid, and $\eta$ is the maximum energy transfer per unit of the incident particle energy.

The closure relationship is provided by Poisson equations,

$$\frac{\partial^2 \phi}{\partial x^2} = -\frac{e}{\varepsilon_0}(n_i - Zn), \tag{5}$$

$$\frac{\partial^2 \psi}{\partial x^2} = 4\pi G nm, \tag{6}$$

where, G is the gravitational constant. The charge and the radius of the grain may be expressed as follows,

$$\frac{q}{q_0} = \frac{a}{a_0}, \tag{7}$$

$$a = \left(\frac{3}{4\pi\rho}\right)^{1/3} m^{1/3}. \tag{8}$$

In this model we aim to single out the sputtering effect only, thus we disregard some aspects of the dusty plasma. We consider cold plasma components, i.e. $T_i \approx 0$ and $T_d \approx 0$, inertialess ions and a constant ion numerical density, i.e. the plasma is coupled to a particle reservoir.

To study the stability of our system that is infinite and homogeneous, we linearize Eqs.(1-8) and get,

$$\frac{e}{m_i}\partial_x f = -\partial_x y, \qquad (9)$$

$$n_0\partial_t m + m_0\partial_t n + n_0 m_0 \partial_x v = -\mathbf{l}\, n_0 m_0^{2/3}\left(\frac{n}{n_0} + \frac{2}{3}\frac{m}{m_0}\right), \qquad (10)$$

$$\partial_t v - \frac{eZ_0}{m_0}\partial_x f = -\partial_x y, \qquad (11)$$

$$\partial_x^2 f = \frac{e}{\mathbf{e}_0}(nZ_0 + n_0 Z), \qquad (12)$$

$$\partial_x^2 y = \mathbf{w}_{Jd}^2\left(\frac{m}{m_0} + \frac{n}{n_0}\right), \qquad (13)$$

$$\frac{Z}{Z_0} = \frac{a}{a_0} = \frac{1}{3}\frac{m}{m_0}. \qquad (14)$$

where, $\mathbf{w}_{Jd}^2 = 4\mathbf{p} G n_0 m_0$ is Jeans frequency. The dispersion relation is obtained after Laplace-Fourier transforming Eqs.(9-14) and performing some algebra,

$$\Omega^2 + \frac{\mathbf{l}\, m_0^{-1/3}}{2\mathbf{x}}\left(\frac{\mathbf{w}_{Jd}^2}{\mathbf{w}_d^2} - \mathbf{x}\right)\Omega - \mathbf{w}_{Jd}^2\frac{1+\mathbf{x}}{\mathbf{x}} = 0, \qquad (15)$$

where, $\mathbf{x} = \frac{m_0}{m_i Z_0}$, $\mathbf{w}_d$ is the dust plasma frequency and $\Omega = i\mathbf{w}$.

We distinguish between two cases corresponding to $\mathbf{x} < \frac{\mathbf{w}_{Jd}^2}{\mathbf{w}_d^2}$ and $\mathbf{x} > \frac{\mathbf{w}_{Jd}^2}{\mathbf{w}_d^2}$. Indeed, for $\mathbf{x} > \frac{\mathbf{w}_{Jd}^2}{\mathbf{w}_d^2}$, the roots of the dispersion relation (15) are,

$$2\Omega_{1,2}^> = \frac{\mathbf{l}\, m_0^{-1/3}}{2\mathbf{x}}\left(\mathbf{x} - \frac{\mathbf{w}_{Jd}^2}{\mathbf{w}_d^2}\right)\left[1 \pm \sqrt{1 + \frac{16\mathbf{w}_{Jd}^2(1+\mathbf{x})\mathbf{x}}{\mathbf{l}^2 m_0^{-2/3}(\mathbf{x} - \frac{\mathbf{w}_{Jd}^2}{\mathbf{w}_d^2})^2}}\right], \qquad (16)$$

whereas for $\mathbf{x} < \frac{\mathbf{w}_{Jd}^2}{\mathbf{w}_d^2}$, the roots are given by,

$$2\Omega^<_{1,2} = \frac{l\, m_0^{-1/3}}{2x}\left(\frac{w_{Jd}^2}{w_d^2} - x\right)\left[-1 \pm \sqrt{1 + \frac{16 w_{Jd}^2 (1+x) x}{l^2 m_0^{-2/3}(x - \frac{w_{Jd}^2}{w_d^2})^2}}\right]. \qquad (17)$$

It appears that the modes $\Omega_1^>$ and $\Omega_1^<$ are zero frequency damped modes while $\Omega_2^>$ and $\Omega_2^<$ are purely growing modes. To implement our model we consider grains of graphite immersed in a hydrogen plasma. The choice of the graphite material is realistic for most grains in astrophysical media are grown from silicate and graphite.[2]

For $n_i = 10^{17} m^{-3}$, $a = 10^{-6} m$, $r = 2200\, kg\, m^{-3}$, $m_t = 12.01\, m_p$, $h \approx 0.29$, $U_0 = 7.2\, eV$, $Y_s \approx 10^{-2}$, one finds $x \gg \frac{w_{Jd}^2}{w_d^2}$. Hence, the instability has the growth rate given by,

$$g_2^> = -\Omega_2^> = \frac{l\, m_0^{-1/3}}{4}\left[\sqrt{1 + \frac{16 w_{Jd}^2 m_0^{2/3}}{l^2}} - 1\right] \approx 1.51 \times 10^{-6}\, rad/s. \qquad (18)$$

The sputtering time turns out to be, $t_{sp} \approx 0.66 \times 10^6 s \approx 7.64\ days$.

The growth rate may be written in the following way as well.

$$w^* g_2^> \approx w_{Jd}^2, \qquad (19)$$

where, $w^* = \frac{l\, m_0^{-1/3}}{2}$.

A more realistic approach may be developed, where ions and electrons that are taken Boltzmannian, do not experience significantly the gravitational effect. Dropping Eq.(9) demands an additional constraint; the conservation of the grains number is then retained. The following linearized equations are added to improve the model, namely,

$$\frac{\partial n}{\partial t} + n_0 \frac{\partial v}{\partial x} = 0, \qquad (20)$$

$$\frac{n_i}{n_{i0}} = -\frac{ef}{T_i}. \qquad (21)$$

$$\frac{n_e}{n_{e0}} = \frac{ef}{T_e}. \qquad (22)$$

Poisson's equation should be modified accordingly,

$$\partial_x^2 f = -\frac{e}{e_0}(n_i - n_e - nZ_0 - n_0 Z). \qquad (23)$$

The normal modes analysis lead to the following dispersion relation,

$$\Omega^3 - \frac{4w^*}{3}\Omega^2 + \left(\frac{w_d^2}{L} - w_{jd}^2\right)\Omega - \frac{2w^*}{3}\left(\frac{w_d^2}{L} + w_{jd}^2\right) = 0, \qquad (24)$$

where, $L = 1 + \frac{1}{k^2 l_D^2}$, $l_D$ being the plasma Debye length.

For $w^* = 0$, that is if we neglect the mass fluctuations, Eq.(24) reduces to Eq.(8) of Ref.(3), namely,

$$\Omega^2 = -\frac{w_d^2}{L} + w_{jd}^2 \qquad (25)$$

The dispersion relation is a cubic equation that may be solved analytically by the Cardan's method[7]. Two solutions of Eq.(25) are retained only, for the third solution corresponds to a negative energy mode that is rejected. These solutions correspond to a damped mode at zero frequency, viz.,

$$w_{r1} = 0, \qquad (26)$$

$$g_1 = \sqrt[3]{b + \sqrt{b^2 + a^3}} - \sqrt[3]{-b + \sqrt{b^2 + a^3}} - \frac{4w^*}{9}, \qquad (27)$$

and a growing mode at frequency rate $w_{r2}$ and a growth rate $g_2$ given by,

$$w_{r2} = \frac{\sqrt{3}}{2}\left\{\sqrt[3]{b + \sqrt{b^2 + a^3}} + \sqrt[3]{-b + \sqrt{b^2 + a^3}}\right\} \qquad (28)$$

$$g_2 = \frac{1}{2}\left\{\sqrt[3]{-b + \sqrt{b^2 + a^3}} - \sqrt[3]{b + \sqrt{b^2 + a^3}}\right\} - \frac{4w^*}{9}, \qquad (29)$$

where, $3a = w_j^2 + \frac{16}{27}w^{*2} - \frac{w_d^2}{L}$, $\frac{b}{w^*} = -\left(\frac{4}{9}\right)^3 w^{*2} - \frac{1}{9}\frac{w_d^2}{L} - \frac{5}{9}w_{jd}^2 < 0$, and $L = 1 + \frac{1}{k^2 l_D^2}$, $l_D$ being the plasma Debye length. It should be noted that, for long wavelengths ($k^2 l_D^2 \ll 1$), $g_2$ is positive provided the condition $w_j^2/w^{*2} > 4.7$ is fulfilled. The normalized growth rate with respect to the sputtering frequency is plotted in Fig.(1). Had we considered the effect of the gravitational potential, we would have found the dispersion relation with a modified Jeans frequency, $\tilde{w}_j^2 = w_j^2\left(1 + \frac{1/x}{k^2 l_i^2 L}\right)$, where $l_i$ is the ion Debye length. However, this contribution is not significant.

To conclude we have investigated the instability of self-gravitating, unmagnetized and collisionless dusty plasmas, where the erosion of dust grains by the ions falling onto their surface is taken into account. The grain mass is then a dynamical variable. A cold plasma approximation is privileged to highlight the sputtering effect. The growth rate of the instability is determined. Two new modes have been derived, viz. a damped mode at zero frequency and a purely growing mode. In an improved model, the electrons that have been dropped earlier are considered Boltzmannian as well as the ions. A cubic dispersion relation is then derived and analytically solved. It is found that two modes, viz., a damped mode at zero frequency and a growing mode at a finite frequency are supported by the plasma. This instability arising due to grain mass fluctuations, may affect the balance between the electrostatic repulsion and the gravitational attraction. It is worthwhile noticing that grain charge fluctuations for a constant mass have been already shown to offset this balance as well. We believe that the sputtering instability affects also, the agglomeration instability as defined in Ref.[6], because it develops on a much longer timescale ( $t_{agg}$ ~ $yrs$ ).

**Acknowledgment**

R.A acknowledges joint financial support of USTHB (Algeria)/UKZN(South-Africa).

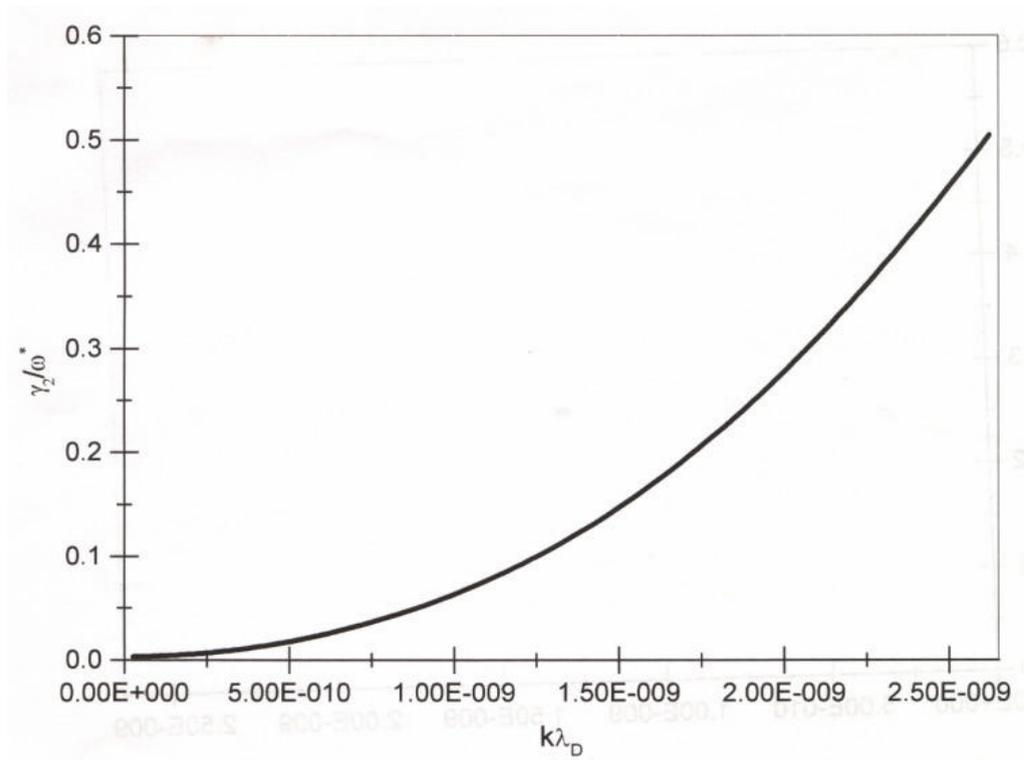

Fig.1 : The normalized growth rate $\gamma_2/\omega^*$ versus $k\lambda_D$, for $\omega_J^2/\omega_d^2 \approx 5$.